\documentclass{aa}  

\usepackage{graphicx}
\usepackage{txfonts}

\usepackage{bm}
\usepackage{xspace}

\newcommand{\planck}{\textsl{Planck}}

\newcommand{\chandra}{\textsl{Chandra}}
\newcommand{\xmm}{\textsl{XMM-Newton}}

\newcommand{\beq}{\begin{equation}}
\newcommand{\eeq}{\end{equation}}
\newcommand{\beqa}{\begin{eqnarray}}
\newcommand{\eeqa}{\end{eqnarray}}
\newcommand{\mpc}{$h^{-1} \mathrm{Mpc}$}
\newcommand{\msun}{$h^{-1} {\rm M}_{\odot}$}
\newcommand{\ms}{${\rm M}_{\odot}$}

\newcommand{\eg}{e.g.,\xspace}

\def\der{{\rm d}}
\usepackage[normalem]{ulem}

\begin{document} 

   \title{Convolutional neural network-reconstructed velocity \\ for kinetic SZ detection}
   
   \titlerunning{CNN-reconstructed velocity for kSZ detection}
   \author{Hideki Tanimura\inst{1} \and Nabila Aghanim\inst{1} \and Victor Bonjean\inst{2,3} \and Saleem Zaroubi\inst{4,5}}
  
   %\subtitle{I. Overviewing the $\kappa$-mechanism}

    \institute{
    Universit\'{e} Paris-Saclay, CNRS, Institut d'Astrophysique Spatiale, B\^atiment 121, 91405 Orsay, France \and 
    Instituto de Astrofísica de Canarias, E-38205 Tenerife, Spain \and 
    University of La Laguna, E-38206 Tenerife, Spain \and
    ARCO (Astrophysics Research Center), Department of Natural Sciences, The Open University of Israel, 1 University Road, PO Box 808, Ra’anana 4353701, Israel \and
    Kapteyn Astronomical Institute, University of Groningen, PO Box 800, NL-9700 AV Groningen, the Netherlands \\
    \email{hideki.tanimura@ias.u-psud.fr}
    }

   \date{}

% \abstract{}{}{}{}{} 
% 5 {} token are mandatory
 
  \abstract 
    {We report the detection of the kinetic Sunyaev-Zel'dovich (kSZ) effect in galaxy clusters with a 4.9$\sigma$ significance using the latest 217~GHz \planck\ map from data release 4. For the detection, we stacked the \planck\ map at the positions of 30 431 galaxy clusters from the Wen-Han-Liu (WHL) catalog. To align the sign of the kSZ signals, the line-of-sight velocities of galaxy clusters were estimated with a machine-learning approach, in which the relation between the galaxy distribution around a cluster and its line-of-sight velocity was trained through a convolutional neural network. To train our network, we used the simulated galaxies and galaxy clusters in the Magneticum cosmological hydrodynamic simulations. The trained model was applied to the large-scale distribution of the Sloan Digital Sky Survey galaxies to derive the line-of-sight velocities of the WHL galaxy clusters. Assuming a standard $\beta$-model for the intracluster medium, we obtained the gas mass fraction in $R_{500}$ to be $f_{gas,500} = 0.09 \pm 0.02$ within the galaxy clusters with the average mass of $M_{500} \sim 1.0 \times 10^{14}$ \msun.}

\maketitle

%
%-------------------------------------------------------------------

\section{Introduction}
\label{sec:intro}

Galaxy clusters are the largest gravitationally bound structures in the Universe and are used as probes of cosmology and astrophysics. These massive objects imprint their signature on the cosmic microwave background (CMB) through the Sunyaev-Zel'dovich (SZ) effect \citep{Sunyaev1970, Sunyaev1972, Sunyaev1980}. The SZ effect is caused by the scattering of CMB photons by hot and ionized plasma in the intracluster medium (ICM), giving rise to a change in the CMB temperature. The SZ effect is classified into two contributions: the thermal Sunyaev-Zel'dovich (tSZ) effect and the kinetic Sunyaev-Zel'dovich (kSZ) effect \citep{Sunyaev1980}.

The kSZ effect is caused by the scattering of CMB photons off the electrons due to the peculiar motion of a galaxy cluster, leading to a Doppler shift of the CMB blackbody spectrum. While the kSZ effect is elusive due to its small amplitude and identical spectral signature to the CMB, it has great potential in constraining both cosmological \citep{Bhattachary2008, Alonso2016, Ma2014, Mueller2015b, Bianchin2016, Madhavacheril2019, Kuruvilla2021a, Kuruvilla2021b} and astrophysical models. 
Although the use of kSZ to constrain cosmological models is still limited due to the weakness of the signal, several studies used the kSZ effect to measure the optical depth or gas mass fraction in galaxy clusters (\eg  \citealt{Planck2016IRXXXVII, Schaan2016, Soergel2016, Bernardis2017, Sugiyama2018, Lim2020, Tanimura2021}). Because the kSZ effect is sensitive to the virialized gas and also to the gas surrounding halos, independent of its temperature (unlike the tSZ effect), it is well suited for studying the gas distribution around galaxy clusters. kSZ can hence potentially be used to solve the debate as to whether a significant fraction of diffuse gas is present around halos as a circumgalactic medium or whether the gas, once expelled because of feedback processes such as star formation, supernovae, and active galactic nuclei (AGNs), is never accreted onto the halos (\eg \citealt{Planck2013IRXI, Anderson2015, Brun2015}). 

The kSZ signal has so far been detected for a few individual systems (\eg \citealt{Sayers2013, Adam2017}) or by statistical measurements such as the pairwise method (\eg \citealt{Hand2012, Hernandez2015, Planck2016IRXXXVII, Soergel2016, Bernardis2017}), cross-correlation method \citep{Hill2016}, and the technique of angular redshift fluctuations \citep{Chaves2021}. In addition to these approaches, an optimized stacking analysis was used by \cite{Schaan2016}, \cite{Schaan2021}, and \cite{Tanimura2021} (hereafter T21).  T21 estimated peculiar velocities of galaxy groups and clusters through the distribution of their surrounding galaxies from the Sloan Digital Sky Survey (SDSS) \citep{Reid2016} based on the linearized continuity equation and used the peculiar velocities to align the signs of the kSZ signals and stack them. While a linearized continuity equation was used to estimate the peculiar velocities in those studies, a deep learning technique was recently used by \cite{Wu2021} to reconstruct the cosmic velocity field from the dark matter density field in numerical simulations. 

In the present study, we extend this approach to real data using the training on galaxy distribution derived from hydrodynamical simulations instead of dark matter distribution. 
We use a new machine-learning approach to estimate the line-of-sight (LOS) velocities of galaxy clusters based on their surrounding galaxy distribution. The purpose is to apply our trained model to actual data and use the estimated LOS velocities to detect the kSZ signal of galaxy clusters. 
The paper is organized as follows. Section \ref{sec:data} summarizes datasets used in our analyses. Section \ref{sec:train} explains the LOS velocity reconstruction of galaxy clusters.  Sections \ref{sec:stack} and \ref{sec:detection} present the stacking method and the detection of the kSZ signals. The interpretation of the measurements is presented in Section \ref{sec:ana}. We end this paper with discussions and conclusions in Section \ref{sec:conclusion}. 

Throughout this work, all masses are quoted in units of solar mass divided by the present value of the Hubble parameter $h$, and $M_{\Delta}$ is mass enclosed within a sphere of radius $R_{\Delta}$ such that the enclosed density is $\Delta$ times the {critical} density at redshift $z$. For the Hubble constant, we used $H_0 = 70.4$ km s$^{-1}$ Mpc$^{-1}$ from \cite{Komatsu2011} in our data analysis, but we obtain consistent results using the value from the \planck\ cosmology in \cite{Planck2016XIII}. Uncertainties are given at the 1$\sigma$ confidence level.

%_______________________________________________________

\section{Data}
\label{sec:data}
%_______________________________________________________

\subsection{Galaxy cluster catalog}
\label{subsec:whl}
A total of 158 103 galaxy groups and clusters (hereafter WHL galaxy clusters) were identified by \cite{Wen2012} and \cite{Wen2015} using the SDSS galaxies in the redshift range between 0.05 and 0.8, of which 89\% have spectroscopic redshifts. The masses of the WHL galaxy clusters were estimated from their total luminosity and were calibrated by the masses of 1191 clusters using X-ray or tSZ measurements. 
From this cluster catalog, T21 selected the WHL galaxy clusters with spectroscopic redshifts at $0.25< z <0.55$ and masses of $M_{500} > 10^{13.5}$\msun. These authors also removed galaxy clusters if their surroundings were largely masked by the Galactic and point-source masks produced by the \textit{Planck} collaboration for their analysis of the CMB and SZ maps (see Sect.\ref{subsec:pr3} for the masks). This selection resulted in the 30 431 galaxy clusters used in T21, which we also use in the present study.  

\subsection{Galaxy catalog}
\label{subsec:sdss}
To estimate the peculiar velocities of WHL galaxy clusters, T21 used the Baryon Oscillation Spectroscopic Survey (BOSS) LOWZ galaxies and constant-mass (CMASS) galaxies in \cite{Reid2016}. This sample is composed of 953 193 galaxies in the northern Galactic hemisphere and 372 542 in the southern Galactic hemisphere. The completeness of the galaxies is stated to be 99\% for CMASS and 97\% for LOWZ. Spectroscopic data are available for all the galaxies, and their redshifts extend up to z $\sim$ 0.8. 
T21 limited their analysis to the redshift range of $0.25 < z < 0.55$, in which the number density of the galaxies in the survey volume is fairly flat (see Fig.~11 in \citealt{Reid2016}). In the present analysis, we also limit to this redshift range. 

\subsection{Planck maps from PR3}
\label{subsec:pr3}
T21 used the \planck\ all-sky map at 217 GHz from the Planck 2018 data release \citep{Planck2018III} for the detection of the kSZ signal\footnote{This map was provided in HEALpix\footnote{http://healpix.sourceforge.net/} format \citep{gorski2005} with a pixel resolution of $N_{\rm side}$ = 2048 ($\sim 1.7$ arcmin).}. This frequency corresponds to the null frequency of the tSZ effect. To minimize the Galactic and extragalactic contamination, T21 applied the mask produced by the \planck\ team for the analysis of the CMB temperature maps, which masks the region around the Galactic plane and the point sources detected at all the \planck\ frequencies (see Table C.1 in \citealt{Planck2018IV}). In addition, T21 used a more robust point-source mask, masking radio and infrared sources used for the analysis of the Compton $y$ maps \citep{Planck2016XXII}. Combining these two masks excludes $\sim$50\% of the sky. We also use these \planck\ maps and masks to compare our results with those of T21. 

\subsection{Planck maps from PR4}
\label{subsec:pr4}
We also apply our new approach on the latest \planck\ PR4 data \citep{Planck2020LVII}\footnote{https://irsa.ipac.caltech.edu/data/Planck/release\_3/ancillary-data/HFI\_Products.html\#hfiallskymaps}.
The \planck\ PR4 data include several improvements compared to the previous data release, namely the use of foreground polarization priors during the calibration stage to break scanning-induced degeneracies, the correction of bandpass mismatch at all frequencies, and the inclusion of 8\% more data collected during repointing maneuvers, and so on.
These improvements reduced noise and systematic effects in the frequency maps at all angular scales and yielded better internal consistency between the various frequency bands. 

\planck\ collaboration split data into two sets and produced two maps at each frequency, which they called half-ring maps. The difference between the two half-ring maps cancels out the astrophysical emissions and can be used as noise maps of the band maps. We use these noise maps for our null tests.

\subsection{Magneticum simulation}
\label{subsec:magneticum}
For the training and test of our machine-learning approach, we use the Magneticum simulation, which is one of the largest cosmological hydrodynamical simulations \citep{Hirschmann2014, Dolag2015}, and is based on the standard $\Lambda$CDM cosmology from \cite{Komatsu2011} with $\Omega_{\rm m} = 0.272$, $\Omega_{\rm b} = 0.046$, and $H_0 = 70.4$ km s$^{-1}$ Mpc$^{-1}$. Several simulation boxes with different sizes and resolutions \footnote{The associated data are available at http://www.magneticum.org/simulations.html} were produced, of which we use ``Box0'' with the largest box size of 2688 \mpc\ using 2 $\times$ 4536$^3$ particles for dark matter and baryons, including the post-processed data of the galaxy catalog and the cluster catalog. In particular, we used the simulation data from ``snapshot 25'' at $z\sim$ 0.47, corresponding to the median redshift of the WHL galaxy clusters used in our analysis. 

%_______________________________________________________

\section{Machine-learning reconstructed LOS velocities}
\label{sec:train}
%_______________________________________________________

\subsection{Training with the Magneticum simulation}
\label{subsec:train}
We trained our network to learn the correlation between the LOS velocities of galaxy clusters and their surrounding galaxies using the Magneticum simulation. The three-dimensional (3D) velocity of galaxy clusters ($\varv_x$, $\varv_y$, and $\varv_z$) is provided by the simulation. We define $\varv_z$ as LOS velocity in the simulation. We note that we did not train for velocities in other directions than LOS because they are irrelevant for our purposes to detect the kSZ signal. 

First, we reproduced the selection performed on the "real" data to construct "mock" data. 
We removed simulated galaxy clusters with $M_{500} < 10^{13.5}$\msun\ (minimum mass of the WHL galaxy clusters in our analysis) and also removed simulated galaxies with $M_{*} < 1.0 \times 10^{10}$ \msun\ so that the number density of galaxies is the same as in the real data. Second, we constructed a grid coordinate system around each galaxy cluster from the simulations as in T21. 
The authors placed a galaxy cluster at the center in a cubic box of $\sim$250$^3$ \mpc, in which the 3D box was divided into grid cells of $5^{3}$ \mpc. Then, T21 placed galaxies around the galaxy cluster in the box cells and calculated the galaxy overdensities. The grid size was determined to be large enough compared to the length expected from redshift-space distortion (RSD): The RSD for a typical velocity of 300 km/s is $\sim$3 \mpc\ at the corresponding redshift range. Box cells were then smoothed by a Gaussian kernel of 2 \mpc\ to remove sharp grid edges.
In our present analysis, the grid size is modified to be 10 \mpc\ because the same grid size as T21 causes a memory issue during our training process. Third, we split the simulation box of 2688 (\mpc)$^3$ into eight independent regions and used seven for the training and validation (the number of galaxy clusters in this volume is 418 374) and one for the test (where the number of galaxy clusters is 59 767). Finally, we input a series of 25$^3$ voxels corresponding to the 3D galaxy overdensity fields around galaxy clusters into our convolutional neural network and train our network for the LOS velocities of the galaxy clusters. 
For comparison, we changed the size of the box from 250$^3$ \mpc\ to 210$^3$ \mpc\ and $\sim$290$^3$ \mpc, but our kSZ measurements were consistent within the 1$\sigma$ uncertainty.

\subsection{Neural network architecture}
\label{subsec:cnn}
We adopt a convolutional neural network (CNN) architecture \citep{cnn}. The overall structure of our network is designed as follows.

A series of 25$^3$ voxels with a volume of 10$^3$ \mpc\ per voxel, corresponding to galaxy overdensity fields with a size of 250$^3$ \mpc, are fed into a CNN with three convolutional layers to capture the abundant features in the 3D fields. Each convolutional layer is designed to have a ``relu'' activation function and ``same'' padding so that the outputs have the same dimension as the inputs. Each layer is then passed to the Maxpooling layer to decrease the dimension by half. This step helps reduce the number of parameters to learn and the number of computations performed in the network. The first, second, and third layers consist of 16, 32, and 64 filters, respectively, each of which has a shape of 3$^3$. We changed the number of filters from {16, 32, and 64} to {32, 64, and 128}, but our kSZ measurements were consistent within the 1$\sigma$ uncertainty. Thus, we used the lower number of filters to reduce the number of parameters to learn and the amount of computations in the network. Finally, the output of the three convolutional layers is fed into two fully connected dense neural layers: the first layer with 64 output units and the second with one output unit, which corresponds to the LOS velocity of a galaxy cluster.  
The learning process is configured with the optimizer and loss function that our model uses. We use ``rmsprop'' for optimizer and ``mse'' for loss function. 

\subsection{Test with the Magneticum simulation}
\label{subsec:test}
\begin{figure*}
    \centering
    \includegraphics[width=0.49\linewidth]{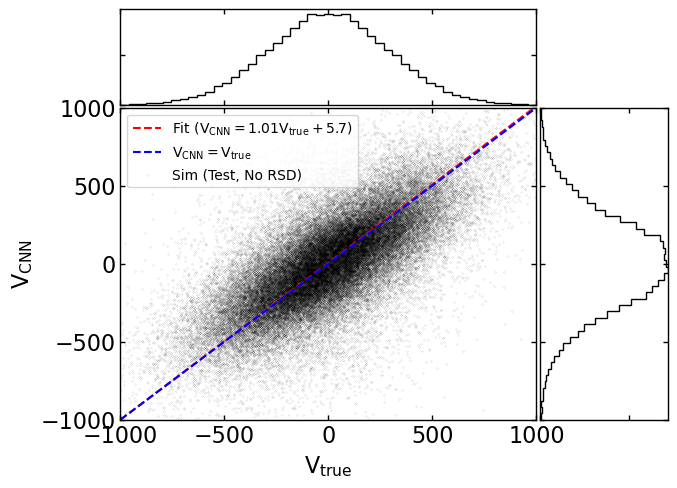}
    \includegraphics[width=0.49\linewidth]{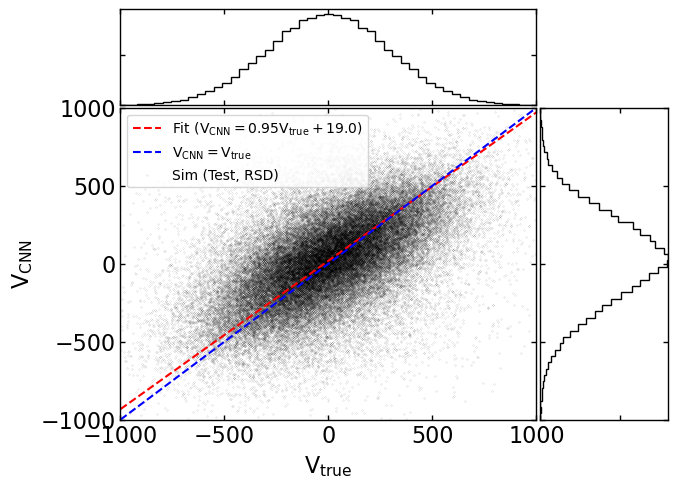}
    \caption{Estimated LOS velocities of simulated galaxy clusters in the Magneticum simulations with our machine-learning approach compared to the true values.
    X-axis: True LOS velocities of simulated galaxy clusters in the Magneticum simulations. Y-axis: Estimated LOS velocities of the same galaxy clusters with our machine-learning approach in Sect.\ref{subsec:train} when the RSD effect is not included  {\it (left)} and included  {\it (right)}. The projected distributions along the X-axis and Y-axis are shown on the top and right panels. In addition, the result of the linear fit is shown between the true and estimated LOS velocities as the red dashed line, which is compared to the case where they are equal, which is shown by the blue dashed line.}
    \label{fig:cnnvel-sim}
    \end{figure*}

\begin{figure*}
    \centering
    \includegraphics[width=0.49\linewidth]{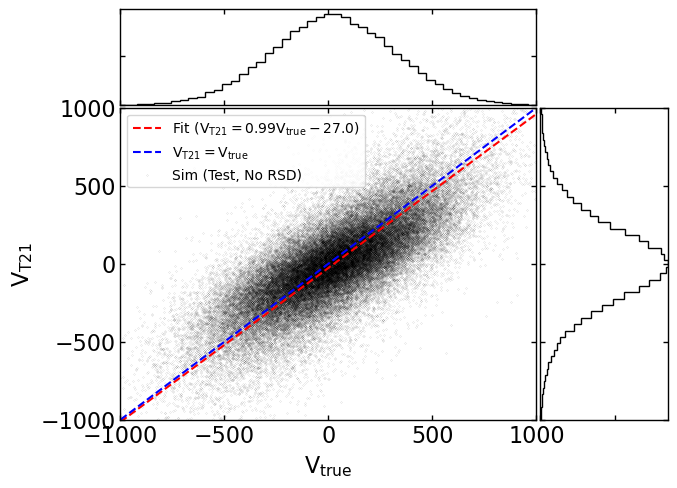}
    \includegraphics[width=0.49\linewidth]{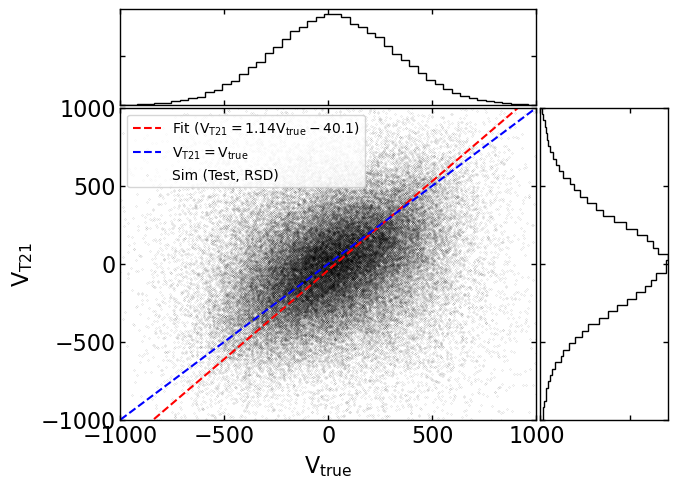}
    \caption{Estimated LOS velocities of simulated galaxy clusters in the Magneticum simulations by the T21's method compared to the true values.
    X-axis: True LOS velocities of simulated galaxy clusters in the Magneticum simulations. Y-axis: Estimated LOS velocities of the same galaxy clusters computed using the method in T21 with the galaxy bias of $b=2$ when the RSD effect is not included  {\it (left)} and included  {\it (right)}. The projected distributions along the X-axis and Y-axis are shown in the top and right panels. In addition, the result of the linear fit is shown between the true and estimated LOS velocities as the red dashed line, which is compared to the case where they are equal, which is shown by the blue dashed line. }
    \label{fig:linvel-sim}
    \end{figure*}

We checked the validity of our machine-learning approach using the test-set in the Magneticum simulation.
The first test was performed with the mock galaxies in the simulations. 
We applied our trained model in Sect.\ref{subsec:train} to the galaxy clusters and their surrounding galaxy overdensity fields in the test-set and estimated their LOS velocities. The result is displayed in the left panel of Fig.~\ref{fig:cnnvel-sim}. The figure shows a positive correlation between the true ($\varv_{\rm true}$) and estimated ($\varv_{\rm CNN}$) LOS velocities. To check the bias in the estimated velocities, we fit them with a linear equation and found little bias with $\varv_{\rm CNN} = 1.01 \times \varv_{\rm true} + 5.7 \,\, \rm [km/s]$. 
The difference between the true and estimated velocities gives the uncertainty associated with our approach, which was estimated to be $\Delta \varv \sim$ 189 km/s. 
For comparison, we used the method in Section 3 of T21 and estimated the LOS velocities of the galaxy clusters ($\varv_{\rm T21}$) in the same test-set.
The result is shown in the left panel of Fig.~\ref{fig:linvel-sim}. Again, to check the bias in the estimated velocities, we fit the true and estimated velocities with a linear equation, and almost no bias was found with $\varv_{\rm T21} = 0.99 \times \varv_{\rm true} -27.0 \,\, \rm [km/s]$. The uncertainty in the estimated velocities was obtained to be $\Delta \varv \sim$ 187 km/s, nearly equivalent to the value derived with our machine-learning approach.

To construct the model for application to the actual data, we then added RSD to the galaxies (Eq. 11 in \citealt{Hogg1999}) in the simulations and re-trained our model. 
We note that we did not re-identify galaxy clusters in the redshift space but used the same galaxy clusters with their positions redshifted.
We applied our re-trained model to the galaxy clusters and their galaxy overdensity fields in the test-set and estimated their LOS velocities. The result is displayed in the right panel of Fig.~\ref{fig:cnnvel-sim}. The estimated LOS velocities show a positive correlation with the true LOS velocities, but give a slightly lower value with $\varv_{\rm CNN} = 0.95 \times \varv_{\rm true} + 19.0 \,\, \rm [km/s]$. The uncertainty in the estimated velocities was obtained to be $\Delta \varv \sim$ 232 km/s.
For comparison, we again used the method in T21 and estimated the LOS velocities of the galaxy clusters in the same test-set. The result is shown in the right panel of Fig.~\ref{fig:linvel-sim}. The estimated LOS velocities show a slightly higher value than the true LOS velocities with $\varv_{\rm T21} = 1.14 \times \varv_{\rm true} -40.1 \,\, \rm [km/s]$. The uncertainty in the estimated velocities was obtained to be $\Delta \varv \sim$ 302 km/s. 
In summary, the bias in the estimated LOS velocities with our machine-learning approach is reduced to $\sim$5\% compared to the bias of $\sim $14\% with the method in T21. In addition, its uncertainty with our machine-learning approach is reduced to $\Delta \varv \sim$ 232 km/s compared to $\Delta \varv \sim$ 302 km/s with the method in T21, indicating an improvement with our machine-learning approach compared to the method in T21.


\subsection{LOS velocities of WHL galaxy clusters}
\label{subsec:whl_vel}

\begin{figure}
    \centering
    \includegraphics[width=1.0\linewidth]{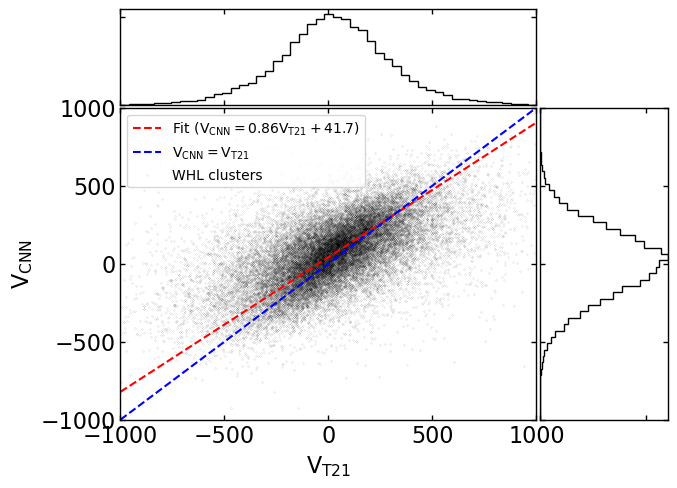}
    \caption{Estimated LOS velocities of the WHL galaxy clusters with our machine-learning approach and the method in T21. The projected distributions along the X-axis and Y-axis are shown in the top and right panels. In addition, the result of the linear fit is shown between the two approaches as the red dashed line, which is compared to the case where they are equal, which is shown by the blue dashed line.}
    \label{fig:whl-vel}
    \end{figure}

We applied our trained model with machine learning in Sect.\ref{subsec:train} to the SDSS galaxy distribution around the WHL galaxy clusters to estimate the LOS velocities of the clusters. We then compared the estimated LOS velocities with the ones computed by the method in T21 in Fig.~\ref{fig:whl-vel}. Our LOS velocities from machine learning are slightly lower by 14\% than those estimated with the approach in T21 with their relation of $\varv_{\rm CNN} \sim 0.86 \times \varv_{\rm T21} +41.7 \,\, \rm [km/s]$. 

This difference is expected from the tests in Sect.\ref{subsec:test} using the test-set in the Magneticum simulations. Our machine-learning approach estimated the LOS velocities to be lower than the true values by $\sim$5\%, and the T21 approach did higher by $\sim$14\%. Thus, the expected difference between these approaches is a factor of 0.95/1.14 $\sim$ 0.83 in the simulations, which is similar to the value of $\sim$ 0.86 in the real data in Fig.~\ref{fig:whl-vel}.

%_______________________________________________________

\section{Stacking analysis}
\label{sec:stack}
%_______________________________________________________

We performed the stacking analysis to detect the kSZ signals from the WHL galaxy clusters using the same method as in T21 but replacing their LOS velocities with those estimated from our machine-learning approach. We stacked the \planck\ map from PR3 used in T21 and also the \planck\ map from the latest PR4. The latest \planck\ data have reduced statistical and systematic uncertainties, which provide more accurate and precise kSZ measurements.

\begin{figure}
    \centering
    \includegraphics[width=\linewidth]{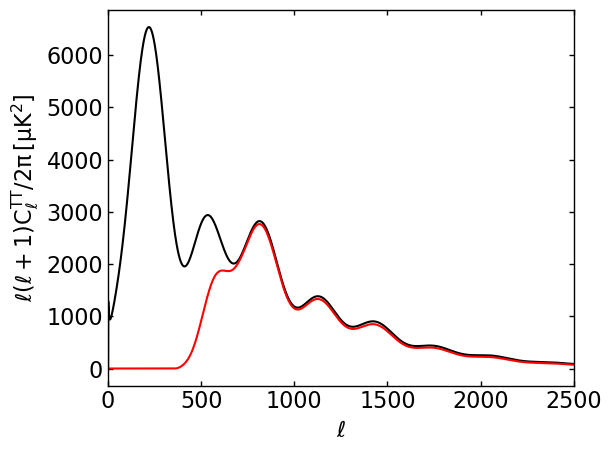}
    \caption{Filter applied to the \planck\ maps. The filter is a smooth function with its response of one below 15 arcmin ($\ell\sim$720) and zero above 30 arcmin ($\ell\sim$360). With this filter, the primordial CMB fluctuations in {\it black} are suppressed to the ones in {\it red}. }
    \label{fig:cmb-filter}
    \end{figure}

Following T21, we first applied the filter shown in Fig.~\ref{fig:cmb-filter}. 
This filter reduces the contaminating contribution from the primordial CMB fluctuations. The amplitude of the primordial CMB fluctuations is on the order of $\sim100$ $\mu$K, while the amplitude of the kSZ signal around galaxy clusters is on the order of $\sim$1 $\mu$K, much weaker than the CMB. Therefore, to increase the signal-to-noise ratio (S/N) of the kSZ signal, we filtered out signals at large scales above 30 arcmin ($\ell\sim$360) that are dominated by the CMB, and kept signals at small angular scales below 15 arcmin ($\ell\sim$720), where the kSZ signal from the WHL galaxy clusters is dominant. (The angular size of the virial radius of the WHL galaxy clusters is 2.2 -- 10.5 arcmin.) With this filter, the standard deviation of the primordial CMB fluctuations is reduced to $\sim40$ $\mu$K. 

Second, we placed each galaxy cluster at the center of a two-dimensional grid in ``scaled'' angular distance in the range of  $-10 < \theta/\theta_{500} < 10$, divided into 10 $\times$ 10 bins, where $\theta_{500}$ is the angular radius of a galaxy cluster calculated with $R_{500}$ provided in the WHL cluster catalog. The \planck\ maps were scaled accordingly, and the data were placed on the two-dimensional grids, while the data in the masked region were not used. 

Third, using the scaled maps, we stacked the data into radial bins, which were also weighted by the LOS velocity and the variance of temperature values within $10 \times \theta_{500}$ for each cluster as follows.
\beq
T(R) = \frac{\sum_i T_i(R) \times  \varv_{i,\rm LOS} / \sigma_i^2}{\sum_i |\varv_{i,\rm LOS}| / \sigma_i^2},
\eeq
where $T_i(R)$ is the temperature value of the $i$-th cluster at the radial distance, $R$, $\varv_{i,\rm LOS}$ is the LOS velocity of the $i$-th cluster, and $\sigma_i$ is the variance of temperature values within the region we consider ($10 \times \theta_{500}$) centered on the $i$-th cluster. 
The weight allows us to align the signs of the kSZ signals given that there is an equal probability that a cluster will have a positive or negative LOS velocity, and the associated kSZ signal from clusters cancels out by a simple stacking \footnote{We define ``positive'' LOS direction as a radial direction from us: a positive motion is a motion moving away from us, and a negative motion is a motion approaching us. It follows that when a galaxy cluster has a positive motion, the CMB is redshifted, resulting in a negative kSZ signal. On the other hand, when a galaxy cluster has a negative motion, the CMB is blueshifted, resulting in a positive kSZ signal.}.
A positive kSZ signal weighted by a negative LOS velocity has a negative signal, and a negative kSZ signal weighted by a positive LOS velocity also has a negative signal. Therefore, the kSZ signals can be stacked without any cancelation, while other components are canceled out. As an additional advantage, a galaxy cluster with a low LOS velocity (i.e., a weak kSZ signal) is underweighted in this stacking.

Finally, the stacked radial profile of the WHL galaxy clusters is computed. We assessed the uncertainties of the stacked profile through bootstrap resampling: We drew a random sample of the 30 431 galaxy clusters with replacement and re-calculated one stacked profile for the new set of 30 431 galaxy clusters. We repeated this process 1000 times and produced the 1000  stacked, bootstrapped profiles with which the covariance between different radial bins was computed. 

%_______________________________________________________

\section{kSZ detection from WHL galaxy clusters}
\label{sec:detection}
%_______________________________________________________

\subsection{kSZ detection with \planck\ PR3 maps}
\label{subsec:pr3ana}
\begin{figure}
    \centering
    \includegraphics[width=1.0\linewidth]{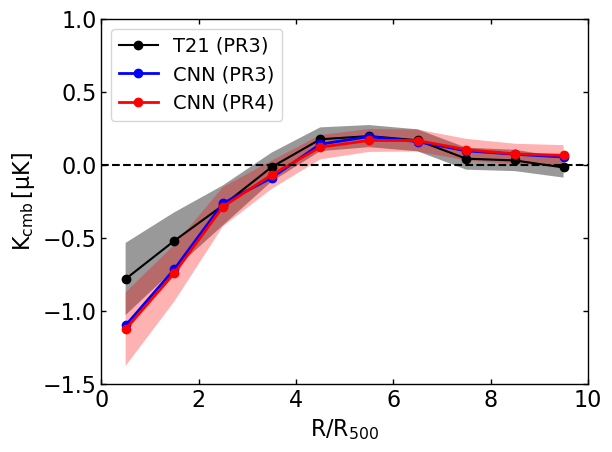}
    \caption{Velocity-weighted kSZ radial profile around the 30 431 WHL galaxy clusters using the CNN-estimated LOS velocities and the \planck\ PR4 temperature map at 217 GHz (CNN PR4, {\it red}). The 1$\sigma$ uncertainty is estimated by a bootstrap resampling. It is compared to the kSZ radial profile using the CNN-estimated LOS velocities and the \planck\ PR3 temperature map at 217 GHz (CNN PR3, {\it blue}). It is also compared to the kSZ radial profile in T21 using the \planck\ PR3 temperature map at 217 GHz (T21 PR3, {\it black}) with the 1$\sigma$ uncertainty in gray. }
    \label{fig:stackp3}
    \end{figure}

We stacked the \planck\ PR3 data at 217 GHz using the LOS velocities from T21 (T21 PR3) and our machine-learning approach (CNN PR3) in Fig.~\ref{fig:stackp3}. As expected, the reduced uncertainties in the CNN-derived velocity estimate induce a better S/N in the kSZ measurements. The oscillating angular pattern seen in the stacked radial profile is the result of the convolution of kSZ and CMB with the filter, as discussed in T21. We refer to this profile as the velocity-weighted kSZ profile from now on. 

We estimated the excess of the measured velocity-weighted kSZ profile with respect to the null hypothesis. The S/N can be estimated as
\beq
S/N = \sqrt{\chi^2_{\rm data} - \chi^2_{\rm null}}
\label{eq:snr}
,\eeq
where
\beqa
\chi^2_{\rm data} &=& \sum_{i,j} T_{\rm data}(R_{i})^{T} (C_{ij}^{-1}) \, T_{\rm data}(R_{j}) ,\\
\chi^2_{\rm null} &=& \sum_{i,j} T_{\rm null}(R_{i})^{T} (C_{ij}^{-1}) \, T_{\rm null}(R_{j}),
\eeqa
where $T_{\rm data}(R_{i})$ is the temperature value at the $R_{i}$ bin of the data kSZ profile, $T_{\rm null}(R_{i})$ is the temperature value at the $R_{i}$ bin under the null hypothesis, which is zero, and $C_{ij}$ is the covariance matrix of the data profile estimated by the bootstrap resampling. By measuring the kSZ signal up to $4\times\theta_{500}$, the S/N was estimated to be $\sim$ 4.7$\sigma$, compared to $\sim$ 3.5$\sigma$ in T21.

\subsection{kSZ detection with \planck\ PR4 maps}
\label{subsec:pr4ana}

\begin{figure*}
    \centering
    \includegraphics[width=0.49\linewidth]{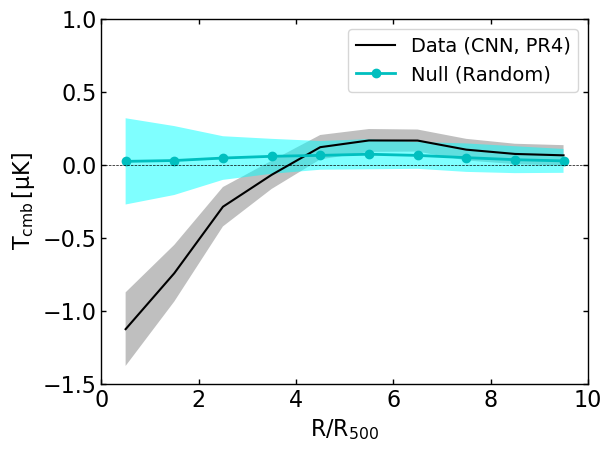}
    \includegraphics[width=0.49\linewidth]{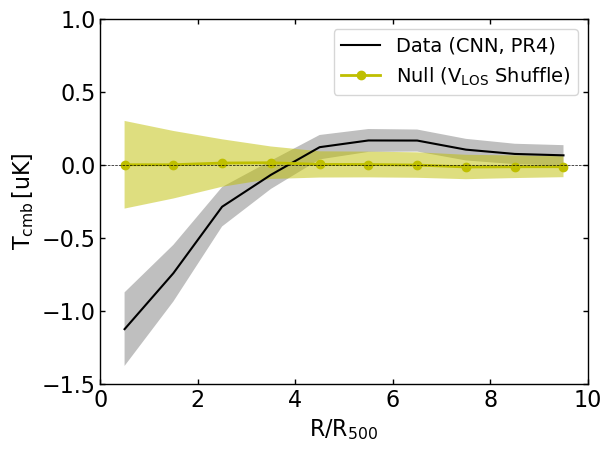}
    \includegraphics[width=0.49\linewidth]{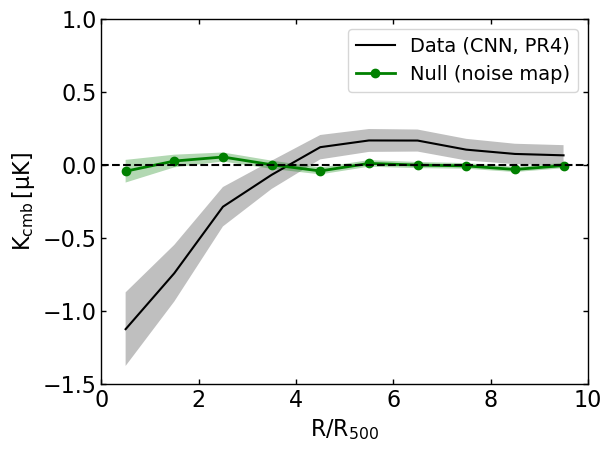}
    \includegraphics[width=0.49\linewidth]{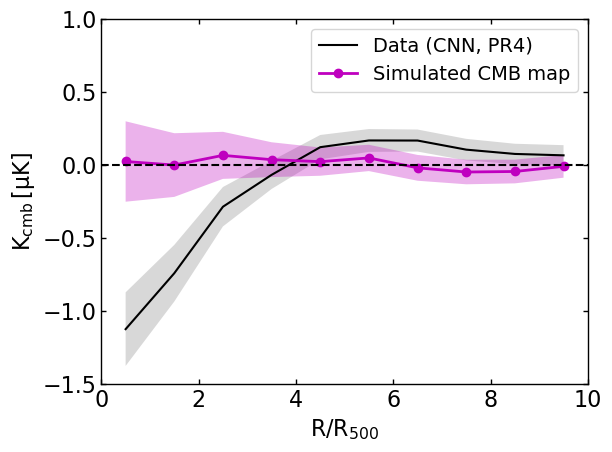}
    \caption{Velocity-weighted kSZ radial profile around the 30 431 WHL galaxy clusters with the \planck\ PR4 temperature map at 217 GHz ({\it black}) compared to the ones from three null tests and a simulated CMB map. In the {\it top left} panel, the clusters are displaced at random positions on the \planck\ map, and then stacked. This process is repeated 1000 times and the mean of the 1000 random samples is computed ({\it cyan}). The 1$\sigma$ uncertainty is estimated by computing a standard deviation of the 1000 random samples. The  {\it top right} panel shows the results from cluster stacking after randomly shuffling the LOS velocities of the clusters; this process is repeated 1000 times and the mean of the 1000 velocity-shuffled profiles is computed ({\it yellow}). The 1$\sigma$ uncertainty is estimated by computing a standard deviation of the 1000 velocity-shuffled profiles. In the {\it bottom left } panel, the clusters are stacked with a noise map produced by $(T_{217}^{\rm HM1} - T_{217}^{\rm HM2})/2$, where $T_{217}^{\rm HM1(2)}$ is the half mission 1(2) \planck\ map at 217 GHz ({\it green}). In the {\it bottom right} panel, the clusters are stacked with a simulated CMB map ({\it magenta}). In both instances, the 1$\sigma$ uncertainty is estimated using a bootstrap resampling. }
    \label{fig:stackp4}
    \end{figure*}

We replaced the \planck\ map from PR3 with the one from the latest PR4 and performed the stacking analysis. The result is shown in Fig.~\ref{fig:stackp4}. The 1$\sigma$ statistical uncertainty is shown as the shaded area, corresponding to the square root of diagonal terms of the covariance matrix estimated by a bootstrap method. 

To confirm the excess, we performed three Monte Carlo-based null tests, as in T21.
In the first null test, we displaced the centers of the galaxy clusters to random positions on the sky and then stacked the \planck\ maps at these random positions. This process is repeated 1000 times to assess the $rms$ fluctuations of the foreground and background signals. This result is shown in {\it cyan} in the top left panel of Fig.~\ref{fig:stackp4} with the rms fluctuations. 
In the second null test, we randomly shuffled the LOS velocities of the galaxy clusters, and then the clusters were stacked with weights based on the shuffled LOS velocities. This process sets the correlation between LOS velocities and clusters to zero. This shuffling process was repeated 1000 times, and we evaluated the mean and standard deviation of the 1000 stacked profiles with shuffled velocities. The result is shown in {\it yellow} in the top right panel of Fig.~\ref{fig:stackp4} with the rms fluctuations.
In the third null test, we stacked with a noise map produced from $(T_{217}^{\rm HM1} - T_{217}^{\rm HM2})/2$, where $T_{217}^{\rm HM1(2)}$ is the half mission 1(2) \planck\ map at 217 GHz. The result is shown in {\it green} in the bottom left panel of Fig.~\ref{fig:stackp4} with the one $\sigma$ uncertainties. This result suggests that the contribution from instrumental noise is minor. 
As a conclusion from the three null tests, the average of the null-test profiles is consistent with zero, suggesting that our measurements are unbiased. 

To check the contribution to the uncertainty from the primordial CMB fluctuations, 
we simulated a primordial CMB fluctuation map based on the WMAP7 cosmological parameters \citep{Komatsu2011}, applied the same filter on the map as for the real data, and performed the stacking analysis. The result is shown in {\it magenta} in the bottom right panel of Fig.~\ref{fig:stackp4} with the one $\sigma$ uncertainties. This result suggests that the uncertainty in our measurement is dominated by the primordial CMB fluctuations.

Finally, we estimated the excess of the measured velocity-weighted kSZ profile to the null hypothesis. The S/N was estimated to be $\sim$ 4.9$\sigma$. 

%_______________________________________________________

\section{Gas mass fraction in WHL galaxy clusters}
\label{sec:ana}
%_______________________________________________________

In this section, we model our kSZ measurements and estimate the gas mass fraction in the WHL galaxy clusters. The relative variation of CMB temperature due to the kSZ effect is given by
\beq
\frac{\Delta{T}_{\rm ksz}}{T_{\rm CMB}} = - \sigma_{\rm T} \int n_{\rm e} \, \left( \frac{\bm{\varv}\cdot \bm{\hat{n}}}{c}  \right) \, \der{l} \simeq - \tau \left( \frac{\bm{\varv} \cdot \bm{\hat{n}}}{c} \right)
\label{eq:ksz}
,\eeq
where $\sigma_{\rm T}$ is the Thomson scattering cross section, $c$ is the speed of light, $n_{\rm e}$ is the electron number density, and $\bm{\varv} \cdot \bm{\hat{n}}$ represents the peculiar velocity of electrons along the line of sight. The integral, $\tau = \sigma_{\rm T} \int n_{\rm e} \der{l} $, is performed assuming that the typical correlation length of LOS velocities (given by $\bm{\varv} \cdot \bm{\hat{n}}$) is much larger than the gas density correlation length ($\sim$5\mpc). Thus, the velocities can be considered to be almost constant. This assumption is justified by \cite{Planck2016IRXXXVII} who showed that the typical correlation length of peculiar velocities is 80–100 \mpc, well above the gas correlation length. 

The density profile in galaxy clusters can be expressed with a $\beta$ model \citep{Cavaliere1978} given by 
\beq
n_{\rm e}(r) = n_{\rm e,0} \left[ 1 + \left( \frac{r}{r_{\rm c}} \right)^2 \right]^{-3\beta / 2}, 
\label{eq:n3d}
\eeq
where $n_{\rm e,0}$ is the central electron density, $r$ is the cluster radial extension, and $r_{\rm c}$ is the core radius of the electron distribution. We used $\beta=0.86$ and $r_{\rm c} = 0.2 \times R_{500}$ from the measurements of the South Pole Telescope clusters \citep{Plagge2010}. The observed profile is given by the geometrical projection of the 3D density profile, which is given by  
\beq
\tau(R) = \sigma_{\rm T} \int \frac{2r \, n_{\rm e}(r)}{\sqrt{r^2 - R^2}} \, \der r, 
\label{eq:n2d}
\eeq
where $R$ is the tangential distance from a galaxy cluster. (We represent the 3D distance with the lowercase letter $r$, and the 2D distance on a map with the uppercase letter $R$.) This 2D projected profile is finally convolved with the angular filter shown in Fig.~\ref{fig:cmb-filter}. 

\begin{figure}
    \centering
    \includegraphics[width=1.0\linewidth]{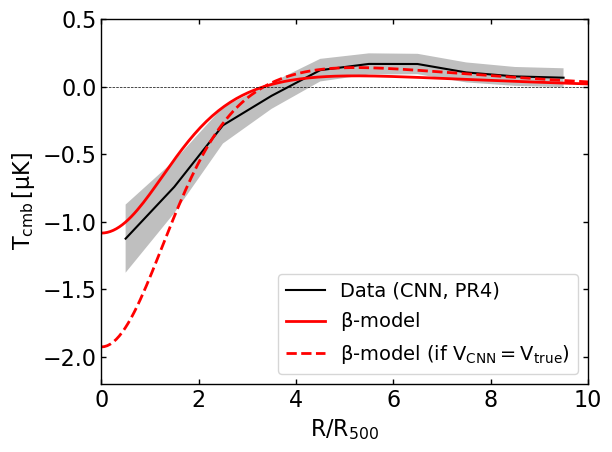}
    \caption{Velocity-weighted kSZ radial profile around the 30 431 WHL galaxy clusters with the \planck\ PR4 temperature map at 217 GHz ({\it black}), fitted with the $\beta$ model ({\it red solid line}). Expected kSZ signal if $\varv_{\rm CNN} = \varv_{\rm true}$ is shown in {\it red dashed line}. }
    \label{fig:modelfit}
    \end{figure}

We fit the $\beta$ model to the measured kSZ profile. For the LOS velocity term in Eq.\,\ref{eq:ksz}, we used the average LOS velocity (in absolute value) of the WHL galaxy clusters from Sect.\ref{subsec:whl_vel}. However, the uncertainties on the LOS velocities induce a decrease in the amplitude of the measured kSZ signal \citep{Nguyen2020} due to the stacks of kSZ signals with the wrong sign. This effect can be analytically corrected using the uncertainty value of the LOS velocities estimated from the Magneticum simulation in Sect. \ref{subsec:test}. Including this correction in the model, the result of the fit is shown in Fig.~\ref{fig:modelfit}. The reduced $\chi^2$ value is 0.6. We note that a coherent angular pattern in the model profile, which is similar to the data profile, is due to the convolution of the $\beta$ profile with our filter.
The optical depth of intracluster gas in the cluster within $R_{500}$ is defined as 
\beq
\tau_{\rm e, 500} = \int_{0}^{R_{500}} \sigma_{\rm T} \, n_{\rm e}(r) \, \der{V}, 
\eeq
and the fit provides an average optical depth of the WHL galaxy clusters of 
\beq
\overline{\tau}_{\rm e, 500} =  (2.4 \pm 0.5) \times 10^{-3}. 
\label{eq:tau-measure}
\eeq

The total gas mass in a galaxy cluster can be defined as  
\beq
M_{gas,500} = \int_{0}^{R_{500}} n_{\rm e}(r) \, \mu_{\rm e} \, m_{\rm p} \, \der{V}, 
\label{eq:mgas}
\eeq
where $\mu_{\rm e} =  1.148$ is the mean molecular weight of electrons \citep{Arnaud2010}, and $m_{\rm p}$ is the mass of proton. From this equation, we can thus compute the average gas mass in the WHL galaxy clusters. It is estimated to be $\overline{M}_{gas,500} \sim 0.9 \times 10^{13}$ \msun. This provides a gas mass fraction of $f_{gas,500} = M_{gas,500}/M_{500} = 0.09 \pm 0.02$ given their average total mass of $M_{500} \sim 1.0 \times 10^{14}$ \msun.

%________________________________________________________
\section{Discussion and Conclusion}
\label{sec:conclusion}
%________________________________________________________

In this paper, we present the detection of the kSZ signal with a significance of $\sim$ 4.9$\sigma$. 
The measurement was performed by stacking the latest \planck\ temperature map at 217 GHz at the positions of the galaxy clusters constructed from the SDSS galaxies. Simple stacks cancel out the kSZ signals because galaxy clusters have an equal probability of having positive or negative kSZ signals. Thus, to avoid this cancelation and align the signs of the kSZ signals, we used the LOS velocities of the WHL galaxy clusters. There is an additional advantage to this approach: other components such as CMB, CIB, and tSZ all have positive signs and are canceled by the stacking approach. The contamination level of these foreground emissions was studied by T21, who showed that it is minor. 

To estimate the LOS velocities of the WHL galaxy clusters, we used a machine-learning approach. In this method, we trained our network to learn the correlation between the LOS velocities of galaxy clusters and their surrounding galaxies with a CNN. To apply our trained model to real data of the WHL galaxy clusters and their surrounding SDSS galaxies, the training was performed on mocks constructed from the Magneticum hydrodynamical simulations. 

There are two main advantages to our new approach of estimating the LOS velocities. 
The LOS velocities of the WHL galaxy clusters were estimated in T21 by relying on the galaxy bias value of the SDSS galaxies derived from other studies \citep{Parejko2013, White2011, Nuza2013, Rodriguez2016}. By contrast, our present model does not need the galaxy bias value, which is implicitly learned by the training process. Moreover, our present model reduces the uncertainty in our LOS velocity estimates due to RSD. These advantages contribute to the increase in the S/N in our kSZ measurements from $\sim$ 3.5 in T21 to $\sim$ 4.7. Furthermore, the S/N is improved to $\sim$ 4.9 by using the latest \planck\ map from the \planck\ 2020 data release.

Based on our new kSZ measurement, we estimated the average optical depth and found $\overline{\tau}_{\rm e, 500} = (2.4 \pm 0.5) \times 10^{-3}$ for galaxy clusters with mass of $M_{500} \sim 1.0 \times 10^{14}$ \msun\ assuming a $\beta$ model. This provides an average gas fraction of $f_{gas,500} = 0.09 \pm 0.02$ within $R_{500}$, which is slightly lower than but consistent with the value in T21 of $f_{gas,500} = 0.12 \pm 0.04$. 
We also compared our value of the gas fraction with that from the Magneticum hydrodynamic simulations. The Magneticum simulation gives $f_{gas,500} \sim 0.13$ for the same mass of galaxy clusters. We also checked with the IllustrisTNG (TNG300-1) \citep{Nelson2019}, giving $f_{gas,500} \sim 0.13$. Our value is slightly lower than the predictions from these hydrodynamic simulations but is again consistent within $\sim$ 2$\sigma$.

Moreover, we compared our value with the measurements from X-rays \citep{Gonzalez2013} ($f_{gas,500} \sim 0.1$ for galaxy clusters with $M_{500} \sim 1.0 \times 10^{14}$ \ms), including \xmm\ and \chandra\ observations from \cite{Vikhlinin2006}, \cite{Sun2009}, and \cite{Sanderson2013}, and kSZ \citep{Soergel2016} ($f_{gas,500} = 0.08 \pm 0.02$ for galaxy clusters with $M_{500} \sim (1-3) \times 10^{14}$ \ms). These results are consistent with our findings. On the other hand, \cite{Lim2020} claims that the gas fraction in halos is approximately equal to the universal baryon fraction down to low-mass halos with $2 \times 10^{12}$ \ms. 
The difference may come from the sample selection; for example, due to the difference in cluster redshifts. \cite{Lim2020} studied a cluster sample at $z$ < 0.12, while our sample and the sample in \cite{Soergel2016} are at $z \sim 0.5$. Therefore, the evolution of the gas in halos may explain the difference in the kSZ measurements. However, the X-ray measurements in \cite{Gonzalez2013} were also applied to local clusters at $z\sim0.1$, and the evolution does not seem to explain the difference. 
Recently, \cite{Schaan2021} detected the kSZ signals at $\sim$6.5$\sigma$ using the Atacama Cosmology Telescope (ACT) DR5 data. Applying our new approach to the ACT data may improve the S/N of the kSZ measurements and help to understand the discrepancy in the gas mass fraction between (\citealt{Lim2020}) and (\citealt{Soergel2016, Gonzalez2013}).
So far, the reason for the discrepancy is unknown, and more precise measurements of gas are necessary in order to make firm conclusions. 

%________________________________________________________
\begin{acknowledgements}
The authors thank an anonymous referee for the useful comments and suggestions. 
This research has been supported by the funding for the ByoPiC project from the European Research Council (ERC) under the European Union's Horizon 2020 research and innovation programme grant agreement ERC-2015-AdG 695561. The authors acknowledge fruitful discussions with the members of the ByoPiC project (https://byopic.eu/team). 
Furthermore, S.Z. acknowledge support by the Israel Science Foundation (grant no. 255/18).
This publication used observations obtained with \planck (\url{http://www.esa.int/Planck}), an ESA science mission with instruments and contributions directly funded by ESA Member States, NASA, and Canada. The authors thank Klaus Dolag and Antonio Ragagnin for providing the Magneticum simulations.
\end{acknowledgements}
%_____________________________________________________________________
\bibliographystyle{aa} % style aa.bst
\bibliography{kszcnn} % your references Yourfile.bib
\end{document}